\documentclass[10pt,twoside]{article}
\usepackage{amsmath}
\usepackage{amssymb}
\usepackage{graphicx}
\usepackage{psfig}
\makeindex
\newcommand{\beq}{\begin{equation}}
\newcommand{\eeq}{\end{equation}}
\newcommand{\beqn}{\begin{eqnarray}}
\newcommand{\eeqn}{\end{eqnarray}}

\newcommand{\bJ}{\mathbf{J}}

\newcommand{\be}{\mbox{${\beta}$}}
\newcommand{\ga}{\mbox{${\gamma}$}}

\newcommand{\de}{\mbox{${\delta}$}}

\begin{document}

\title{Quantized Black Holes, Correspondence Principle, and Holographic Bound}

\markboth{I.B.\,Khriplovich}{Quantized Black Holes}

\author{I.B.\,Khriplovich
\\[5mm]
\it Budker Institute of Nuclear Physics, 630090 Novosibirsk,
Russia,\\ \it and Novosibirsk University\\ \it e-mail:
khriplovich@inp.nsk.su}

\date{}
\maketitle

\thispagestyle{empty}

\begin{abstract}
\noindent An equidistant spectrum of the horizon area of a
quantized black hole does not follow from the correspondence
principle or from general statistical arguments. Such a spectrum
obtained earlier in loop quantum gravity (LQG) does not comply
with the holographic bound. This bound fixes the Barbero-Immirzi
parameter of LQG, and thus leads to the unique result for the
spectrum of horizon area. \\

\end{abstract}

\section{Is Spectrum of Quantized Black Hole\\ Equidistant?}
The idea of quantizing the horizon area of black holes was
put forward many years ago by Bekenstein in the pioneering
article~\cite{bek}. It was based on the intriguing observation,
made by Christodoulou and Ruffini~\cite{ch,chr}: the horizon area
of a nonextremal black hole behaves as an adiabatic invariant. Of
course, the quantization of an adiabatic invariant is perfectly
natural, in accordance with the correspondence principle.

One more conjecture made in~\cite{bek} is that the spectrum of a
quantized horizon area is equidistant. The argument therein was
that a periodic system is quantized by equating its adiabatic
invariant to $2\pi \hbar n$, $n=0,\, 1\, 2\, ...$.

Later it was pointed out by Bekenstein~\cite{be1} that the
classical adiabatic invariance does not guarantee by itself the
equidistance of the spectrum, at least because any function of an
adiabatic invariant is itself an adiabatic invariant. However, up
to now articles on the subject abound in assertions that the form
\beq\label{eq}
A = \be\, l_p^2 n, \quad n= 1,\, 2,\, ...
\eeq
for the horizon area spectrum \footnote{Here and below $l_p^2
=\hbar k/c^3$ is the Planck length squared, $l_p = 1.6 \cdot
10^{-33}$ cm, $k$ is the Newton gravitational constant; $\be$ is
here some numerical factor.} is dictated by the respectable
correspondence principle. The list of these references is too
lengthy to be presented here.

Let us consider an instructive example of the situation when a
nonequidistant spectrum arises in spite of the classical adiabatic
invariance. We start with a classical spherical top of an angular
momentum $\bJ$. Of course, the projection $J_z$ of $\bJ$ is an
adiabatic invariant. If the $z$-axis is chosen along $\bJ$, the
value of $J_z$ is maximum, $J$, or $\hbar j$ in the quantum case.
The classical angular momentum squared $J^2$ is also an adiabatic
invariant, with eigenvalues $\hbar^2 j(j+1)$ when quantized. Let
us try now to use the operator $\hat{J}^2$ for the area
quantization in quite natural units of $l^2_p$. For the horizon
area $A$ to be finite in the classical limit, the power of the
quantum number $j$ in the result for $j \gg 1$ should be the same
as that of $\hbar$ in $l^2_p$~\cite{kh}. With $l^2_p \sim \hbar$,
we arrive in this way at
\[
A \sim l^2_p \sqrt{j(j+1)}.
\]
Since $\sqrt{j(j+1)} \to j+ 1/2$ for $j \gg 1$, we have come back
again to the equidistant spectrum in the classical limit. However,
the equidistance can be avoided in the following way. Let us
assume that the horizon area consists of sites with area on the
order of $l^2_p$, and ascribe to each site $i$ its own quantum
number $j_i$ and the contribution $\sqrt{j_i(j_i+1)}$ to the area.
Then the above formula changes to
\beq\label{sp}
A \sim l^2_p \sum_i \sqrt{j_i(j_i+1)}
\eeq
(in fact, this formula for a quantized area arises as a special
case in loop quantum gravity, see below). Of course, to retain a
finite classical limit for $A$, we should require that $\sum_i
\sqrt{j_i(j_i+1)} \gg 1$. However, any of $j_i$ can be well
comparable with unity. So, in spite of the adiabatic invariance of
$A$, its quantum spectrum (\ref{sp}) is not equidistant, though of
course discrete.

One more argument in favour of the equidistant spectrum (\ref{eq})
is as follows\cite{be1,bem,bemu}. On the one hand, the entropy $S$
of a horizon is related to its area $A$ through the
Bekenstein-Hawking relation
\beq\label{BH}
S = \frac{A}{4 l_p^2}\,.
\eeq
On the other hand, the entropy is nothing but $\ln g(n)$ where the
statistical weight $g(n)$ of any state $n$ is an integer.
In~\cite{be1,bem, bemu} the requirement of integer $g(n)$ is taken
literally, and results after simple reasoning not only in
equidistant spectrum (\ref{eq}), but also in the following allowed
values for the numerical factor $\be$ in this spectrum:
\[
\be = 4 \ln k, \quad k=2,\, 3, \, ... .
\]

Let us imagine however that with some model for $S$ one obtains
for $g(n)$, instead of an integer value $K$, a noninteger one
$K+\de\,,\;\; $ $0 < \de <1$. Then, the entropy will be
\[
S = \ln(K+\de) = \ln K + \de/K.
\]
Now, the typical value of the black hole entropy $S = \ln K = A/4
l_p^2$ is huge, something like $10^{76}$. So, the correction
$\de/K$ is absolutely negligible as compared to $S = \ln K$.
Moreover, it is far below any conceivable accuracy of a
description of entropy in general, and of the semiclassical
Bekenstein-Hawking formula (\ref{BH}) relating the black hole
entropy to the quantized horizon area in particular. Therefore,
this correction can be safely omitted and forgotten. As usual for
macroscopic objects, the fact that $g(n)$ is an integer has no
consequences for entropy.

Thus, contrary to the popular belief, the equidistance of the
spectrum for the horizon area does not follow from the
correspondence principle and/or from general statistical arguments
\cite{khr}.

\section{``It from Bit'', Loop Quantum Gravity,\\ and Holographic Bound}
It does not mean however that any model leading to an equidistant
spectrum for the quantized horizon area should be automatically
abandoned. Quite simple and elegant version of such a model, so
called ``it from bit'', for a Schwarzschild black hole was
formulated by Wheeler~\cite{whe}. The assumption is that the
horizon surface consists  of $\nu$ sites, each of them supplied
with an ``angular momentum'' quantum number $j$ with two possible
projections $\pm 1/2$. The total number $K$ of degenerate quantum
states of this system is obviously
\beq\label{kifb}
K=2^\nu.
\eeq
Then the entropy of the black hole is
\beq\label{S1/2}
S_{1/2}= \ln K= \nu \ln 2.
\eeq
And finally, with the Bekenstein-Hawking relation (\ref{BH}) one
obtains for the area spectrum the following equidistant formula:
\beq\label{A1/2}
A_{1/2} = 4 \ln 2 \;l_p^2 \;\nu.
\eeq
This model of the quantized Schwarzschild black hole looks quite
OK by itself.

Later this result was derived in Ref.~\cite{asht} in the framework
of loop quantum gravity (LQG)~[11-15]. We discuss below whether
the ``it from bit'' picture, if considered as a special case of
the area quantization in LQG, can be reconciled with the
holographic bound~[16-18].

More generally, a quantized surface in LQG is described as
follows. One ascribes to it a set of punctures. Each puncture is
supplied with two integer or half-integer ``angular momenta''
$j^u$ and $j^d$:
\beq\label{j}
j^u,\, j^d= 0, 1/2, 1, 3/2, ...\; ;
\eeq
at least one of them should not vanish. $j^u$ and $j^d$ are
related to edges directed up and down the normal to the surface,
respectively, and add up into an angular momentum $j^{ud}$:
\beq\label{ud}
{\bf j}^{ud}= {\bf j}^{u} + {\bf j}^{d}; \quad |j^{u}-j^{d}|\leq
j^{ud} \leq j^{u}+j^{d}.
\eeq
The area of a surface is
\beq\label{Aj}
A =8\pi\ga\, l_p^2 \sum_i \sqrt{2 j^u_i(j^u_i+1)+ 2j^d_i(j^d_i+1)-
j^{ud}_i(j^{ud}_i+1)}\;.
\eeq
The numerical factor $\ga$ in (\ref{Aj}) cannot be determined
without an additional physical input. This ambiguity originates
from a free (so called Barbero-Immirzi) parameter \cite{imm,rot}
which corresponds to a family of inequivalent quantum theories,
all of them being viable without such an input. Once the general
structure of the horizon area in LQG is fixed, the only problem
left is to determine this overall factor, i. e., the
Barbero-Immirzi parameter.

To make the discussion more clear and concrete, we confine from
now on to the simplified version (quite popular now) of general
formula (\ref{Aj}) when $j^{d(u)}=0$, so that this formula for a
surface area reduces to
\beq\label{A1}
A =8\pi\ga\, l_p^2 \sum_i \sqrt{j_i(j_i+1)}\,, \quad j=j^{u(d)}.
\eeq
It is worth mentioning that this particular case of general
formula (\ref{Aj}) coincides with the naive model (\ref{sp}).

The result (\ref{A1/2}) was obtained in~\cite{asht} under an
additional condition that the gravitational field on the horizon
is described by the $U(1)$ Chern-Simons theory \footnote{Earlier
attempts to calculate the black hole entropy in LQG go back at
least to Refs. \cite{rov,kra}. They did not lead however to
concrete quantitative results.}. Formula (\ref{A1/2}) is a special
case of general one (\ref{A1}) when all $j=1/2$. As to the overall
factor $\ga$, its value here is
\beq\label{ga}
\ga = \frac{\ln 2}{\pi \sqrt{3}}\,.
\eeq

Let us turn now to the holographic bound~[16-18]. According to it,
the entropy $S$ of any spherical nonrotating system confined
inside a sphere of area $A$ is bounded as follows:
\beq\label{hb}
S \leq \frac{A}{4 l_p^2},
\eeq
with the equality attained only for a system that is a black hole.

A simple intuitive argument confirming this bound is as
follows~\cite{sus}. Let us allow the discussed system to collapse
into a black hole. During the collapse the entropy increases from
$S$ to $S_{bh}$, and the resulting horizon area $A_{bh}$ is
certainly smaller than the initial confining one $A$. Now, with
the account for the Bekenstein-Hawking relation (\ref{BH}) for a
black hole, we arrive through the obvious chain of (in)equalities
\[
S \leq S_{bh} = \frac{A_{bh}}{4 l_p^2} \leq \frac{A}{4 l_p^2}
\]
at the discussed bound (\ref{hb}).

The result (\ref{hb}) can be formulated otherwise. Among the
spherical surfaces of a given area, it is the surface of a black
hole horizon that has the largest entropy.

The last statement was used as an assumption by Vaz and
Witten~\cite{vaz} in a model of the quantum black hole as
originating from a dust collapse. Then this assumption was
employed by us~\cite{kk} in the problem of quantizing the horizon
of a black hole in LQG. In particular, the coefficient $\ga$ was
calculated in Ref.~\cite{kk} in the case when the area of a black
hole horizon is given by the general formula (\ref{Aj}) of LQG, as
well as under some more special assumptions on the values of
$j^u$, $j^d$, $j^{ud}$. Moreover, it was demonstrated in \cite{kk}
for a rather general class of the horizon quantization schemes
that it is the maximum entropy of a quantized surface that is
proportional to its area.

\section{Problem of Distinguishability of Edges}
We consider in fact the ``microcanonical'' entropy $S$ of a
quantized surface defined as the logarithm of the number of states
of this surface for a fixed area $A$ (instead of fixed energy in
common problems). Obviously, this number of states $K$ depends on
the assumption made on the distinguishability of the edges. So,
let us discuss first of all which of {\it a priori} possible
assumptions is reasonable from the physical point of view
\cite{khr}.

It is instructive to start the discussion with a somewhat more
detailed analysis of the ``it from bit'' model. The natural result
(\ref{kifb}) for the total number of states in this model
corresponds in fact to the assumption that among all the sites
with $j=1/2$, those with same $j_z$, $\;+1/2$ or $\;-1/2$ (their
numbers are $\nu_+$ and $\nu_-$, respectively), are
indistinguishable, and those with different projections are
distinguishable. Indeed, in this case the number of states with
given $\nu_+$, $\nu_-$ is obviously $\nu!/(\nu_+!\,\nu_-!)$, and
the total number of states and entropy,
\beq\label{sk}
K = \sum_{\nu_+, \; \nu_-}\,\frac{\nu!}{\nu_+!\,\nu_-!}\,=\,
\sum_{\nu_+ = 0}^{\nu}\,\frac{\nu!}{\nu_+!\,(\nu - \nu_+)!}\,=
2^\nu \quad {\rm and} \quad S=\nu \ln 2\,,
\eeq
coincide with (\ref{kifb}) and {\ref{S1/2}).

It should be noted that the configuration with $\nu_+ = \nu_- =
\nu/2$ not only gives here the maximum contribution to $K$ and
$S$. For $\nu \gg 1$ both the ``it from bit'' statistical weight
and entropy are strongly dominated by this configuration
\cite{khri}. Indeed,
\beq\label{skm}
\ln \frac{\nu!}{(\nu/2!)^2}\,= \nu\ln2 -\,\frac{1}{2}\,\ln \nu.
\eeq

Let us also note in passing that if we confine here to the states
with the total angular momentum $J=0$, the entropy is still about
the same:
\[
\nu\ln2 -\,\frac{3}{2}\,\ln \nu.
\]

Now, we are going over to the analysis of the more general case
with various $j$'s.

One possibility, which might look quite appealing here, is that of
complete indistinguishability of edges. It means that no
permutation of any edges results in new states. To analyze this
situation, let us rewrite formula (\ref{A1}) as follows:
\beq\label{A2}
A = 8\pi\ga\, l_p^2 \sum_{jm}\sqrt{j(j+1)}\;\nu_{jm}\,.
\eeq
Here the projections $m$ of an angular momentum $j$ run as usual
from $-j$ to $j$; $\nu_{jm}$ is the number of edges with given $j$
and $m$. Under the assumption of complete indistinguishability,
the total number of angular momentum states created by $\nu_j=
\sum_m \nu_{jm}$ edges of a given $j$ with all $2j+1$ projections
allowed, is \footnote{Perhaps, the simplest derivation of this
formula is as follows. We are looking here effectively for the
number of ways of distributing $\nu_j$ identical balls into $2j+1$
boxes. Then, the line of reasoning presented in \cite{lls}, \S 54,
results in formula (\ref{id}). I am grateful to V.F. Dmitriev for
bringing to my attention that formula (\ref{id}) can be derived in
this simpleminded way.}
\beq\label{id}
K_j = \,\frac{(\nu_j+2j)!}{\nu_j!\; (2j)!}\,.
\eeq
The partial contributions $s_j=\ln K_j$ to the black hole entropy
$S=\sum_j s_j$ that can potentially dominate the numerically large
entropy, may correspond to the three cases: $j \ll \nu_j\,$, $j
\gg \nu_j\,$, and $j \sim \nu_j \gg 1$. These contributions are as
follows:\\

\begin{tabular}[h]{ll}
\vspace{3mm} $j \ll \nu_j\,$, & $s_j \approx 2j\ln \nu_j\,$; \\
 \vspace{3mm}
$j \gg \nu_j\,$,  & $s_j \approx \nu_j\ln j$; \\
 \vspace{3mm}
$j \sim \nu_j \gg 1$,  & $s_j \sim 4j\ln 2$.\\
\end{tabular}\\
In all the three cases the partial contributions to the entropy
$S$ are much smaller parametrically than the corresponding
contributions
\\

 \hspace{3mm} $a_j \sim j \nu_j$\\

\noindent to the area $A=\sum_j a_j$. Thus, in all these cases $S
\ll A$, so that with indistinguishable edges of the same $j$, one
cannot make the entropy of a black hole proportional to its area
\cite{aps,khri}.

Let us consider now the opposite assumption, that of completely
distinguishable edges. In this case the total number of states is
$K = \nu\,!$, with the microcanonical entropy $S=\nu \ln \nu$. In
principle, this entropy can be made proportional to the black hole
area $A$. The model (though not looking natural) could be as
follows. We choose a large quantum number $J  \gg 1$, and assume
that the horizon area $A$ is saturated by the edges with $j$ in
the interval $J~<j~<2J$, and with ``occupation numbers'' $\nu_j
\sim \ln J$. Then, the estimates both for $S$ and $A$ are $\sim
J\ln J$, and the proportionality between the entropy and the area
can be attained.

However, though under the assumption of complete
distinguishability the entropy can be proportional to the area,
the {\it maximum} entropy for a given area is much larger than the
area itself. Obviously, here the maximum entropy for fixed $A \sim
\sum_j \sqrt{j(j+1)}\; \nu_j$ is attained with all $j$'s being as
small as possible, say, $1/2$ or 1. Then, in the classical limit
$\nu \gg 1$, the entropy of a black hole grows faster than its
area: while $A \sim \nu$, $S = \nu \ln \nu \sim A\ln A$. Thus, the
assumption of complete distinguishability is in conflict with the
holographic bound, and therefore should be discarded
\footnote{There is no disagreement between this our conclusion and
that of Refs. \cite{aps,pol,gs}: what is called complete
distinguishability therein, corresponds to another assumption,
last one in this section.}.

Let us go over to the third conceivable possibility, which is
quite popular (see, for instance, Ref. \cite{aps}). Here the
assumption is that the total number of states is
\beq\label{pro}
K=\prod_j (2j+1)^{\nu_j},
\eeq
with the entropy of the horizon surface
\beq\label{prod}
S=\sum_j \nu_j \ln(2j+1).
\eeq
Obviously, in this case the maximum entropy $S_{max} \sim A$ is
reached with the smallest possible value of $j$ for each edge.
Thus, one arrives here effectively to the ``it from bit'' picture
which may look attractive.

However, it can be easily demonstrated that this scheme
corresponds to the following assumption on the distinguishability
of edges:\\

\begin{tabular}[h]{cccc}
\vspace{3mm}
 nonequal $j$, & any $m$ & $\longrightarrow$ &
indistinguishable;\\
\vspace{3mm} equal $j$, & nonequal $m$ &
$\longrightarrow$ & distinguishable;\\
 \vspace{3mm}
equal $j$, & equal $m$ & $\longrightarrow$ & indistinguishable.\\
\end{tabular}\\

\noindent Comparison of the first two lines in this table
demonstrates that this assumption looks strange and unnatural.

Thus, the only reasonable assumption on the distinguishability of
edges that may result in acceptable physical predictions (i. e.,
may comply both with the Bekenstein-Hawking relation and with the
holographic bound) is as follows:\\

\begin{tabular}[h]{cccc}
 \vspace{3mm}
 nonequal $j$, & any $m$ & $\longrightarrow$ &
distinguishable;\\ \vspace{3mm} equal $j$, & nonequal $m$ &
$\longrightarrow$ & distinguishable;\\
 \vspace{3mm}
equal $j$, & equal $m$ & $\longrightarrow$ & indistinguishable.\\
\end{tabular}\\

\section{Microcanonical Entropy of Black Hole}
Under the last assumption, the number of states of the horizon
surface for a given number $\nu_{jm}$ of edges with momenta $j$
and their projections $j_z=m$, is obviously
\beq\label{mk}
K = \nu\,!\, \prod_{jm}\,\frac{1}{\nu_{jm}\,!}\;, \quad \nu
=\sum_j \nu_j = \sum_{jm} \nu_{jm}\,,
\eeq
and the corresponding entropy equals
\beq\label{ms}
S=\ln K = \ln(\nu\,!)\,- \sum_{jm}\,\ln(\nu_{jm}\,!)\,.
\eeq
The structures of the last expression and of formula (\ref{A2})
are so different that in a general case the entropy certainly
cannot be proportional to the area. However, this is the case for
the maximum entropy in the classical limit.

In the classical limit, with all effective ``occupation numbers''
large, $\nu_{jm} \gg 1$, the entropy in the Stirling approximation
is
\beq\label{en2}
S= \sum_{jm} \nu_{jm} \times \ln \left(\sum_{j^{\prime}m^{\prime}}
\nu_{j^{\prime}m^{\prime}}\right)-\sum_{jm} \nu_{jm} \,\ln
\nu_{jm}\,.
\eeq
We calculate its maximum for a fixed area $A$, i. e., for a fixed
sum
\beq\label{N}
N\,=  \sum_{jm}^\infty \sqrt{j(j+1)}\,\nu_{jm}={\rm const} \,.
\eeq

The problem reduces to the solution of the system of equations
\beq\label{sys}
\ln \left(\sum_{j^{\,\prime}m^{\prime}}
\nu_{j^{\,\prime}m^{\prime}}\right) - \ln \nu_{jm} = \mu
\sqrt{j(j+1)}\,,
\eeq
or
\beq\label{nu}
\nu_j = (2j+1)\, e^{- \mu \sqrt{j(j+1)}}\,\nu \,.
\eeq
Here $\mu$ is the Lagrange multiplier for the constraining
relation (\ref{N}). Summing expressions (\ref{nu}) over $j$, and
recalling that $\sum_j \nu_j = \nu$, we arrive at the equation for
$\mu$:
\beq\label{equ}
\sum_{j=1/2}^\infty (2j+1)\, e^{- \mu \sqrt{j(j+1)}}= 1,
\eeq
with the solution
\beq
\mu = 1.722.
\eeq

On the other hand, when multiplying equation (\ref{sys}) by
$\nu_{jm}$ and summing over $jm$, we arrive with the constraint
(\ref{N}) at the following result for the maximum entropy for a
given value of $N$:
\beq\label{enf}
S_{\rm max}= 1.722\,N\,.
\eeq
Now, with the Bekenstein-Hawking relation we find the value of the
Barbero-Immirzi parameter:
\beq\label{bip}
\ga = 0.274.
\eeq

The above derivation follows closely Ref. \cite{kk} (see also
\cite{khr}).

It should be emphasized that this calculation is not special for
LQG only, but applies (with obvious modifications) to a more
general class of approaches to the quantization of surfaces. The
following assumption is really necessary here: the surface should
consist of sites of different sorts, so that there are $\nu_i$
sites of each sort $i$, with a generalized effective quantum
number $r_i$ (here $\sqrt{j(j+1)}$), and a statistical weight
$g_i$ (here $2j+1$). Then in the classical limit, the maximum
entropy of a surface can be found, at least numerically, and it is
certainly proportional to the area of the surface.

Few more comments are appropriate here.

A nice feature of the obtained picture is that here the occupation
numbers $\nu_j$ have a sort of Boltzmann distribution (see
(\ref{nu})), where the partial contributions $\sqrt{j(j+1)}$ to
the horizon area correspond to energies, and $\mu$ is the analogue
of the inverse temperature.

The next point worth mentioning here is as follows. By
substituting the occupation numbers given by formula (\ref{nu})
into expression (\ref{N}), one arrives at the conclusion that both
the effective quantum number $N$ and with it the horizon area $A$
are proportional to an integer $\nu$. It may create the impression
of the equidistant area spectrum for a black hole. However, one
should keep in mind that relation (\ref{nu}) for occupation
numbers is an approximate one, it is valid only for $\nu_j \gg 1$
in the leading approximation to the Stirling formula. In fact, all
$\nu_j$'s are integers, and therefore the exact formula (\ref{N})
in no way corresponds to an equidistant spectrum.

And at last, let us note that the leading correction to our
simplest version (\ref{en2}) of the Stirling formula results in
the correction $\sim \ln^2 A$ to the Bekenstein-Hawking relation
(\ref{BH}) \cite{kk}.

\section{Again Loop Quantum Gravity\\ and Holographic Bound}
We come back now to the result of Ref.~\cite{asht}. If one assumes
that the value (\ref{ga}) of the parameter $\ga$ is the universal
one (i. e., it is not special to black holes, but refers to any
quantized spherical surface), then the value (\ref{S1/2}) is not
the maximum one in LQG for a surface of the area (\ref{A1/2}).
This looks quite natural: with the transition from the unique
choice made in \cite{asht}, $j^{u(d)}=1/2$, $j^{d(u)}=0$, to more
extended and rich one, the number of quantum states should,
generally speaking, increase. And together with this number, its
logarithm, which is the entropy of a quantized surface, increases
as well.

To prove this statement we rewrite formula (\ref{S1/2}) as
follows:
\beq\label{S'}
S_{1/2}=\, \ln 2\,\sqrt{\frac{4}{3}}\,N\, = \,0.800\, N\,, \quad
N= \sqrt{\frac{3}{4}}\,\nu\,,
\eeq
and consider this value of $N$, together with the horizon area
$A$, as a fixed one. This is certainly less than the above result
(\ref{enf}) $1.722\,N$.

As expected, in the general case, with $N$ given by formula
(\ref{N}) with all values of $j^u_i$, $j^d_i$, $j^{ud}_i$ allowed
and $g_i= 2j^{ud}_i +1$, the maximum entropy is even
larger~\cite{kk}
\beq\label{s3}
S_{\rm max}=\,3.120\,N.
\eeq
The corresponding value of the Barbero-Immirzi parameter in this
case is
\beq\label{bip1}
\ga = 0.497.
\eeq

Thus, the conflict is obvious between the holographic bound and
the result (\ref{S'}) advocated in \cite{asht}.

One might try to avoid the conflict by assuming that the value
(\ref{ga}) for the Immirzi parameter $\ga$ is special for black
holes only, while for other quantized surfaces $\ga$ is smaller.
However, such a way out would be unattractive and unnatural.

Quite recently the result of \cite{asht} was also criticized and
revised in \cite{lew,mei}. It is instructive to compare the
starting points of Refs. \cite{lew,mei} with ours. What both
approaches have in common, is the LQG expression (\ref{A1}) for
the quantized horizon area $A$. But in other respects the
difference between them is drastic.

Our approach is based on the natural and transparent physical
requirement that the ``occupation numbers'' $\nu_{jm}$ (see
formulae (19)--(22)) should guarantee the maximum black hole
entropy $S$ for given horizon area $A$. Our additional assumption,
formulated in section 3, is also natural and physically sound:
only those edges that have same $jm$ are indistinguishable. Then
the problem reduces to a straightforward and simple calculation
presented in section 4.

On the other hand, the approach of Refs. \cite{lew,mei} is based
on the equidistance assumption, going back to \cite{ack,abk} and
formulated in \cite{lew} as follows: ``the fixed classical area
$a$ is quantized in the following way,
\beq\label{a}
a = 4\pi\ga l_p^2\, k\,, \quad k \in \mathbb{N}\,,
\eeq
where $k$ is arbitrary.'' However, as has been demonstrated
already in section 1, there are no physical arguments in favor of
an equidistant spectrum. Thus, this assumption is in fact an {\it
ad hoc} one.

Nevertheless, the equation derived in \cite{mei} for the
Barbero-Immirzi parameter (rewritten in our notations)
\beq\label{equ1}
2 \sum_{j=1/2}^\infty e^{- 2 \pi \gamma \sqrt{j(j+1)}}= 1
\eeq
is rather close to ours (\ref{equ}) ($\mu$ in (\ref{equ}) is equal
to $2 \pi \ga$). But I do not see any reasons why the number of
states for a given $j$ should be here 2, instead of the usual
$2j+1$~\footnote{The comment on formula (\ref{equ1}) in \cite{mei}
(also rewritten in our notations) is: ``... if the number of
states for a given spin $j$ was $2j+1$ instead of 2 one would have
in (\ref{equ1}) $2j+1$ instead of 2 in front of the exponential
and the appropriate $\ga$ would be equal to 0.273985635...''.}.
Still, since the sum is strongly dominated by the first term, with
$j=1/2$, the result 0.238 for $\ga$, obtained in \cite{mei}, is
close to ours 0.274.

The conclusions can be summarized as follows.

\noindent 1. The equidistant result of Ref.~\cite{asht} is not
true.

\noindent 2. The value of the Barbero-Immirzi parameter, $\ga =
0.274$, and thus the spectrum of a quantized horizon in LQG (with
expression (\ref{A1}) for area), are fixed uniquely by the
holographic bound, i. e., by the requirement that among the
spherical surfaces of a given area, it is the horizon surface that
has the maximum entropy.

\section*{Acknowledgements}
I am grateful to V.F. Dmitriev, V.M. Khatsymovsky, and G.Yu. Ruban
for discussions. The investigation was supported in part by the
Russian Foundation for Basic Research through Grant No.
03-02-17612.

\end{document}